\documentclass[english,tightenlines,eqsecnum,amsmath,amssymb,aps,prd,superscriptaddress,nofootinbib,showpacs,floats]{article}

\usepackage{epsfig}
\usepackage{babel}
\usepackage{authblk}
\usepackage{amsmath,amsfonts,amssymb,multicol}
\usepackage[usenames]{xcolor}
\definecolor{AHZ}{rgb}{0.0,0.0,0.9}
\definecolor{AHZ1}{rgb}{1,0.0,0.1}
\usepackage[colorlinks,linkcolor=AHZ,citecolor=AHZ1,bookmarks]{hyperref}
\long\def\/*#1*/{}
\usepackage{color}

{\definecolor{RED}{rgb}{1,0,0}
\definecolor{GREEN}{rgb}{0,1,0}
\definecolor{BLUE}{rgb}{0,0,1}
\usepackage{graphicx,caption,subcaption}

\begin{document}

\title{Consequences of energy conservation violation: Late time solutions of $\Lambda({\sf T}) {\sf CDM}$ subclass of $f({\sf R},{\sf T})$ gravity using dynamical system approach}
\author[1]{Hamid Shabani,\thanks{h.shabani@phys.usb.ac.ir}}
\author[2]{Amir Hadi Ziaie,\thanks{ah.ziaie@gmail.com}}

\affil[1]{Physics Department, Faculty of Sciences, University of Sistan and Baluchestan, Zahedan, Iran}
\affil[2]{Department of Physics, Kahnooj Branch, Islamic Azad University, Kerman, Iran}

\date{\today}
%
\maketitle
\begin{abstract}
\noindent
Very recently, the authors of [PRL {\bf 118} (2017) 021102] have shown that violation of energy-momentum tensor ({\sf EMT}) could result in an accelerated expansion state via appearing an effective cosmological 
constant, in the context of unimodular gravity. Inspired by this outcome, in this paper
we investigate cosmological consequences of violation of the {\sf EMT} conservation 
in a particular class of $f({\sf R},{\sf T})$ gravity when only the pressure-less fluid is 
present. In this respect, we focus on the late 
time solutions of models of the type $f({\sf R},{\sf T})={\sf R}+\beta \Lambda(-{\sf T})$.
As the first task, we study the solutions when the conservation of {\sf EMT} is respected and then we proceed with those in which violation occurs. We have found, provided that the {\sf EMT} conservation is 
violated, there generally exist two accelerated expansion solutions which their 
stability properties depend on the underlying model. More exactly, we 
obtain a dark energy solution for which the effective equation of state ({\sf EoS}) depend on model parameters and a de Sitter solution. We present a method to parametrize $\Lambda(-{\sf T})$ function which is useful in dynamical 
system approach and has been employed in the herein model. Also, we discuss the cosmological solutions for
models with $\Lambda(-{\sf T})=8\pi G(-{\sf T})^{\alpha}$ in the presence of the ultra relativistic matter. 
\end{abstract}
%
\section{Introduction}\label{intro}
Today\rq{}s  astrophysical measurements reveal that the Universe is experiencing an accelerated expansion phase~\cite{supno1,supno2,supno3,WMAP1,Teg1,SDSS1,SDSS2,WMAP2,WMAP3,WMAP4,WMAP5}. These set of observational
data have driven the quest for strong theoretical explanations of such a phenomenon. Among the various proposed models, the most popular one is the theory of general relativity ({\sf GR}) modified by a cosmological constant term $\Lambda$, which is called \lq\lq{}the  concordance\rq\rq{} or $\Lambda${\sf CDM} model~\cite{LCDM}. In this model, it is assumed that the $\Lambda$ term may take over the recent eras of the dynamical evolution of the Universe after domination of what is called \lq\lq{}Dark Matter\rq\rq{} ({\sf DM}) which its interactions is still somewhat obscure.
Observational data have discovered that at least $70\%$ of the total energy budget of the Universe is in the 
form of  the so-called \lq\lq{}Dark Energy\rq\rq{} ({\sf DE}) which is regarded as a cosmic medium with unusual properties attributed to cosmological constant effects. These data show that the $\Lambda${\sf CDM} model is in a good accommodation~\cite{sn1,sn2,suzuki}. In spite of its fine agreements with the observation data, there are two major concerns in this context; the first one is referred to as \lq\lq{}the cosmological constant problem\rq\rq{} which
opens a question about the origin and the great disagreement between theoretical and expected values of the cosmological
constant~\cite{Cos.pro1,Cos.pro2,Cos.pro3}. The other problem deals with this puzzlement that why we happen to live in a 
special era of evolution of cosmos where the contribution of $\Lambda$, {\sf DM} and the baryonic matter are of the same order?
This is pointed out as ``cosmic coincidence problem" in the literature.

These issues have motived people to seek for some other theoretical foundations or at least apply some 
modifications to the assumed $\Lambda${\sf CDM} model. In this respect, cosmological scenarios with running $\Lambda$
have been proposed. The first developments, in this context, have been made by Shapiro 
et al.~\cite{shapiro1,shapiro2,shapiro3,shapiro4}. They have shown that 
there are no sturdy evidences to indicate that the cosmological constant is running or not. This fact, 
encourages one to investigate  cosmological scenarios within different theoretical backgrounds that admit running cosmological parameters. Up to now, different running cosmological constant models have been proposed, among which we can quote: a time
dependent cosmological constant motivated by quantum field theory~\cite{shapiro4,bonanno,urbanowski}, 
a running vacuum in the context of supergravity~\cite{mavromatos}, $\Lambda$(t) cosmology induced by Elko 
fields~\cite{pereira}, running cosmological constant via covariant/non-covariant 
parametrization~\cite{stachowski} and some others~\cite{alcaniz,lima,socorro}.

In this paper we work on a particular subclass of $f({\sf R},{\sf T})$ gravity, in which ${\sf R}$ and ${\sf T}$ are the Ricci scalar
and the trace of {\sf EMT}, respectively. Firstly, this model introduced by Harko, et al.~\cite{fRT1} and
later has been widely investigated in~\cite{fRT2,fRT3,fRT4,fRT5,fRT6,fRT7,fRT8,fRT9,fRT10,fRT11,fRT12,fRT13,fRT14,fRT15,fRT16,fRT17,fRT18,fRT19,fRT20}.
In the background of $f({\sf R},{\sf T})$ gravity, we have investigated a linear combination of the Ricci scalar and an arbitrary 
function of the trace of {\sf EMT}, i.e., $f({\sf R},{\sf T})={\sf R}+\Lambda({\sf T})$. In this model, the Einstein gravity has been modified by a minimally coupled \lq\lq{}trace-dependent\rq\rq{} cosmological constant. One may find some efforts to elaborate cosmological features of $\Lambda({\sf T})${\sf CDM} in the literature.
The idea of a running cosmological constant as $\Lambda({\sf T})$, probably dates back to the paper of Poplawski~\cite{poplawski}. He found that 
$\Lambda({\sf T})$ gravity will reduce to Palatini $f({\sf R})$ gravity when the pressure of fluid is neglected. Besides, he concluded that cosmological data
are consistent with $\Lambda({\sf T})$ gravity without any knowledge about the functionality of $\Lambda({\sf T})$ parameter. In~\cite{ahmed}, Bianchi Type-V cosmological solutions have been derived\footnote{However, it seems that the authors of~\cite{ahmed} have made
a wrong assumption about the conservation of {\sf EMT}. For more details, one can compare the paragraph above equation (6) in the original paper
with equation (2.1) in~\cite{comm}.}. The locally rotationally symmetric (LRS) Bianchi type-I cosmological models have been considered in~\cite{sahoo}. In the most performed works on $\Lambda({\sf T})$ gravity, the {\sf EMT} is forced to be conserved. With this assumption, the authors of~\cite{fRT7,fRT10} have shown that these type of models lead to an accelerated expansion era with an undesirable present value for the {\sf EoS} parameter. 

Up to now, cosmological consequences of violation of the {\sf EMT} conservation have not been studied 
properly. It may be a good idea to consider the minimally coupled part of the $f({\sf R},{\sf T})$ model as a running cosmological constant and inspect its cosmological solutions. As we shall see, when the conservation of {\sf EMT} is allowed to be violated, it would result in a {\sf DE} era which is accompanied by an observationally allowed present values for the {\sf EoS} parameter. Specifically, such a model mimics a de Sitter solution. Interestingly, a similar {\sf DE} solution has been reported in~\cite{josset} which arises from the violation of {\sf EMT} conservation. The authors of~\cite{josset} have pointed out that 
violation of {\sf EMT} conservation can be predicted by modified quantum mechanical models or by models that utilize the causal set approach to 
quantum gravity\footnote{In~\cite{josset}, authors have obtained an effective cosmological constant in the context of
unimodular gravity}.  In our analysis, we  have applied the dynamical system approach and employed a useful method  to parametrize the $\Lambda({\sf T})$ function. In this method we have presented a way to determine whether or not a given model lead to a stable late time solution.

The present paper has been organized as follows: In Section~\ref{Fieldequations}, the field equation of $f({\sf R},{\sf T})$ gravity has been 
reviewed. Besides, the relevant dimensionless parameters which will be used in the construction of subsequent equations 
are introduced. A discussion on the {\sf EMT} conservation will be given as well. In Section~\ref{sec-conserved}, we discuss the late time solutions of the only conserved $f({\sf R},{\sf T})$ model. In this case, some issues have already been illustrated, however, we present some other features. Section~\ref{sec-nconserved-tot}, is devoted to describe the mentioned method. In Section~\ref{sec-nconserved}, we consider models in which the $\Lambda({\sf T})$ function obeys a power law behavior. These models will be investigated independently. Finally, in Section~\ref{conclusion} we summarize our results.
%
\section{Field equations of $f({\sf R},{\sf T})$ gravity}\label{Fieldequations}
In this section, we present the field equations of $f({\sf R},{\sf T})$ modified gravity ({\sf MG}) and discuss the conservation of
{\sf EMT}. We assume that a pressure-less fluid ({\sf DM} along with a baryonic matter) and 
an ultra relativistic matter are present. The action of $f({\sf R},{\sf T})$ gravity
can then be written as
\begin{align}\label{action}
S=\int \sqrt{-g} d^{4} x \left[\frac{1}{2\kappa^{2}} f\Big{(}{\sf R}, {\sf T}^{\textrm{(p, u)}}\Big{)}
+{\sf L}^{\textrm{(total)}} \right],
\end{align}
where we have defined the Lagrangian of the total matter as
\begin{align}\label{lagrangian}
{\sf L}^{\textrm{(total)}}\equiv {\sf L}^{\textrm{(p)}}+{\sf L}^{\textrm{(u)}}.
\end{align}
In the above definitions, we have used ${\sf R}$ and ${\sf T}^{\textrm{(p, u)}}\equiv g^{\mu \nu}
{\sf T}^{\textrm{(p, u)}}_{\mu \nu}$ as the Ricci curvature scalar and the trace of {\sf EMT} of pressure-less and ultra relativistic 
fluids (which we get these matters as the total matter content of the Universe), respectively. The letters \lq\lq{}${\rm p}$\rq\rq{} and \lq\lq{}${\rm u}$\rq\rq{} indicate the pressure-less and ultra relativistic fluids and $g$ is the determinant of the metric, $\kappa^{2}\equiv8 \pi G$ is the gravitational coupling constant and we set $c=1$. Since the ultra relativistic fluid has a 
traceless {\sf EMT} i.e., ${\sf T}^{({\rm u})}=0$, 
we have ${\sf T}^{\textrm{(p, u)}}={\sf T}^{\textrm{(p)}}+{\sf T}^{\textrm{(u)}}={\sf T}^{\textrm{(p)}}\equiv {\sf T}$. 
Hereupon, we will drop the letter \lq\lq{}{\rm p}\rq\rq{} from the trace of pressure-less matter for 
simplicity\footnote{Note that, in the current formulation of $f({\sf R},{\sf T})$ gravity, the 
presence of an ultra relativistic fluid does not affect the results in the sense that only the trace of 
pressure-less fluid couples to the Ricci curvature.}. The {\sf EMT} ${\sf T}_{\mu \nu}^{\textrm{(p, u)}}$ is defined as 
the Euler-Lagrange expression of the Lagrangian of the total matter, i.e.,
\begin{align}\label{Euler-Lagrange}
{\sf T}_{\mu \nu}^{\textrm{(p, u)}}\equiv-\frac{2}{\sqrt{-g}}
\frac{\delta\left[\sqrt{-g}({\sf L}^{\textrm{(p)}}+{\sf L}^{\textrm{(u)}})
\right]}{\delta g^{\mu \nu}}.
\end{align}
The field equations for $f({\sf R},{\sf T})$
gravity can be obtained as~\cite{fRT1}
\begin{align}\label{fRT field equations}
&F({\sf R},{\sf T}) {\sf R}_{\mu \nu}-\frac{1}{2} f({\sf R},{\sf T}) g_{\mu \nu}+\Big{(} g_{\mu \nu}
\square -\triangledown_{\mu} \triangledown_{\nu}\Big{)}F({\sf R},{\sf T})=\nonumber\\
&\Big{(}8\pi G-{\mathcal F}({\sf R},{\sf T})\Big{)}{\sf T}_{\mu \nu}^{\textrm{(p, u)}}-\mathcal {F}({\sf R},{\sf T})\mathbf
{\Theta_{\mu \nu}^{\textrm{(p, u)}}},
\end{align}
where
\begin{align}\label{theta}
\mathbf{\Theta_{\mu \nu}^{\textrm{(p, u)}}}\equiv g^{\alpha \beta}\frac{\delta
{\sf T}_{\alpha \beta}^{\textrm{(p, u)}}}{\delta g^{\mu \nu}},
\end{align}
and for the sake of convenience, we have defined the following
functions as derivatives with respect
to the trace ${\sf T}$ and the Ricci curvature scalar ${\sf R}$
\begin{align}\label{f definitions1}
{\mathcal F}({\sf R},{\sf T}) \equiv \frac{\partial f({\sf R},{\sf T})}{\partial {\sf T}}~~~~~
~~~~~\mbox{and}~~~~~~~~~~
F({\sf R},{\sf T}) \equiv \frac{\partial f({\sf R},{\sf T})}{\partial {\sf R}}.
\end{align}
We consider a spatially flat, homogeneous and isotropic Universe which is described by the Friedmann-Lema\^{\i}tre-Robertson-Walker ({\sf FLRW}) metric as
\begin{align}\label{metricFRW}
ds^{2}=-dt^{2}+a^{2}(t) \Big{(}dr^{2}+r^{2}d\Omega^2\Big{)},
\end{align}
where, $a(t)$ denotes the scale factor of the Universe. Therefore, field equation (\ref{fRT field equations}) by assuming metric (\ref{metricFRW}), leads to
\begin{align}\label{first}
&3H^{2}F({\sf R},{\sf T})+\frac{1}{2} \Big{(}f({\sf R},{\sf T})-F({\sf R},{\sf T}){\sf R}\Big{)}+3\dot{F}({\sf R},{\sf T})H=\nonumber\\
&\Big{(}8 \pi G +{\mathcal F} ({\sf R},{\sf T})\Big{)}\rho^{\textrm{(p)}}+8 \pi G\rho^{\textrm{(u)}},
\end{align}
as the modified Friedmann equation, and
\begin{align}\label{second}
&2F({\sf R},{\sf T}) \dot{H}+\ddot{F} ({\sf R},{\sf T})-\dot{F} ({\sf R},{\sf T}) H=\nonumber\\
&-\Big{(}8 \pi G+{\mathcal F} ({\sf R},{\sf T})\Big{)}\rho^{\textrm{(p)}}-\frac{32}{3} \pi
G\rho^{\textrm{(u)}},
\end{align}
as the modified Raychaudhuri equation. In equations (\ref{first}) and (\ref{second}), $H$ indicates the Hubble
 parameter. Hereafter, we work on the Lagrangians which include minimal coupling between the trace of {\sf EMT} and the Ricci scalar, i.e.,
\begin{align}\label{minimal}
f({\sf R},{\sf T})=g({\sf R})+\beta \Lambda (-{\sf T}),
\end{align}
where $\beta$ can control the strength of the coupling. Since, for the pressure-less matter we 
have ${\sf T}^{\rm{(p)}}=-\rho^{\rm{(p)}}$, in order to avoid ambiguity due to the negative sign, 
we work hereafter with $\Lambda (-{\sf T})$ function. In fact, to guarantee that $\Lambda({\sf T})$ is always 
a real-valued function, we consider $\Lambda (-{\sf T})$ instead. For this class of $f({\sf R},{\sf T})$ models we have
 $\mathcal{F}({\sf R}, {\sf T})=-\beta \Lambda'(-{\sf T})$\footnote{Note that hereafter, we shall 
use ${\mathcal F} ({\sf R},{\sf T})=\partial f({\sf R},{\sf T})/\partial {\sf T}=\beta d\Lambda(-{\sf T})/d{\sf T}=-\beta d\Lambda(-{\sf T})/d(-{\sf T})=-\beta \Lambda'$.} and $F({\sf R}, {\sf T})=g'({\sf R})$. The field equations (\ref{first}) and (\ref{second}) can then be rewritten as
\begin{align}\label{eom1}
&1+\frac{g}{6H^{2} g'} +\beta\frac{\Lambda}{6 H^{2} g'}-\frac{{\sf R}}{6 H^{2}} + \frac{\dot{g'}}
{H g'}=\\\nonumber
&\frac{8 \pi G \rho^{\textrm{(p)}}}{3H^{2} g'}-\beta\frac{\Lambda' \rho^{\textrm{(p)}}}
{3H^{2} g'}+\frac{8 \pi G \rho^{\textrm{(u)}}}{3H^{2} g'},
\end{align}
and
\begin{align}\label{eom2}
&2\frac{\dot{H}}{H^{2}}+\frac{\ddot{g'}}{H^{2} g'} -\frac{\dot{g'}}{H g'}=\\\nonumber
&-\frac{8\pi G \rho^{\textrm{(p)}}}{H^{2} g'}+\beta\frac{\Lambda' \rho^{\textrm{(p)}}}{H^{2} g'}
-\frac{32\pi G \rho^{\textrm{(u)}}}{3H^{2} g'},
\end{align} 
where the arguments have been dropped for convenience. In order to construct the dynamical system 
for the field equations (\ref{eom1}) and (\ref{eom2}), it is helpful to define a few 
dimensionless variables and parameters as
\begin{align}
&x_{1}\equiv-\frac{\dot{g}({\sf R})}{H g({\sf R})},\label{varx1}\\
&x_{2}\equiv-\frac{g({\sf R})}{6 H^{2} g'({\sf R})},\label{varx2}\\
&x_{3}\equiv \frac{{\sf R}}{6 H^{2}}=\frac{\dot{H}}{H^{2}}+2,\label{varx3}\\
&x_{4}\equiv \frac{\kappa^{2} \Lambda (-{\sf T})}{3H^{2} g({\sf R})},\label{varx4}\\
&x_{5}\equiv  -\frac{\kappa^{2}{\sf T} \Lambda'(-{\sf T})}{3H^{2} g({\sf R})},\label{varx5}\\
&\Omega^{\textrm{(u)}}\equiv\frac{\kappa^{2} \rho^{\textrm{(rad)}}}{3 H^{2} g({\sf R})},\label{omega rad}\\
&\Omega^{\textrm{(p)}}\equiv\frac{\kappa^{2} \rho^{\textrm{(p)}}}{3 H^{2} g({\sf R})},\label{omega mat}
\end{align}
where we have used Ricci scalar, ${\sf R}=6(\dot{H}+2H^2)$ for metric (\ref{metricFRW}), within the 
definition of ($\ref{varx3}$). Moreover, we use the following definitions
\begin{align}
&n \equiv -\frac{{\sf T} \Lambda''(-{\sf T})}{\Lambda' (-{\sf T})}\label{n},\\
&s \equiv -\frac{{\sf T}\Lambda'(-{\sf T})}{\Lambda (-{\sf T})}=\frac{x_{5}}{x_{4}}.\label{parameters}
\end{align}
In general, eliminating ${\sf T}$ from (\ref{n}) and (\ref{parameters}) yields 
$n=n(s)$. Describing the models with $n=n(s)$ instead of $\Lambda(-{\sf T})$, 
can be suitable in dynamical system analysis. In the following subsections we discuss the consequences of conservation/non-conservation of {\sf EMT}, which in turn, results in a key equation that helps us to study the dynamical evolution of the models. We classify minimally coupled models as those that respect {\sf EMT} conservation and those that do not.
%
\subsection{Models which obey the conservation of {\sf EMT}}\label{conservedmodel}
In this subsection, we present the results that come from considering the {\sf EMT} conservation. If we apply the Bianchi identity to the field equation (\ref{fRT field equations}) and assume that the conservation of {\sf EMT} holds for pressure-less and ultra relativistic fluids, independently, we get
\begin{align}
&\dot{\rho}^{(\textrm{p})}+3H\rho^{(\textrm{p})}=0,\label{pcon}\\
&\dot{\rho}^{(\textrm{u})}+4H\rho^{(\textrm{u})}=0.\label{ucon}
\end{align}
We then find the following constraint for the pressure-less fluid as\footnote{See \cite{fRT7,fRT10} for more details.}
\begin{align}\label{constraint2}
\dot{h}'=\frac{3}{2}Hh'.
\end{align}
After a straightforward but lengthy algebra, we arrive at a specific form for $\Lambda (-{\sf T})$, as follows
\begin{align}\label{specific}
\Lambda (-{\sf T})=C_{1}\sqrt{-{\sf T}}+C_{2},
\end{align}
where $C_{1}$ and $C_{2}$ are constants of integration. It means that solution (\ref{specific}) is the only subclass of  $f({\sf R},{\sf T})$ theories of gravity with minimal coupling that respect the conservation of {\sf EMT}. For this solution we obtain $x_{5}=x_{4}/2$ which reduces the space constructed from the variables of the theory.
\subsection{Models which violate the conservation of {\sf EMT}}\label{non-conservedmodel}
These models do not generally respect conservation laws (\ref{pcon}) and (\ref{ucon}). Applying the Bianchi identity
to the field equation (\ref{fRT field equations}) leads to the following equation between the function $\mathcal {F}({\sf R},{\sf T})$, the {\sf EMT} and its trace as
\begin{align}\label{relation}
&\kappa^{2}\bigtriangledown^{\mu}{\sf T}^{\rm (u)}_{\mu \nu}+(\kappa^{2} +\mathcal {F}^{\rm (p)})\bigtriangledown^{\mu}{\sf T}^{\rm (p)}_{\mu \nu}+\frac{1}{2}\mathcal {F}^{\rm (p)}\bigtriangledown_{\mu}{\sf T}^{\rm (p)}\nonumber\\
&+{\sf T}^{\rm (p)}_{\mu \nu}\bigtriangledown^{\mu}\mathcal {F}^{\rm (p)}-\nabla_{\nu}(p^{\rm (p)}\mathcal{F}^{\rm (p)})=0,
\end{align}
where the argument of $\mathcal {F}({\sf R},{\sf T})$ has been dropped for abbreviation. Notice that in the above equation, we have used $\mathcal {F}^{\rm (p)}$ in the corresponding terms of pressure-less fluid, since only ${\sf T}^{{\rm (p)}}$ would appear in the argument of $\mathcal {F}({\sf R},{\sf T})$; we further note that the function $\mathcal {F}$ and its derivative are zero for the ultra relativistic fluids. Equation (\ref{relation}) has a specific form such that we can consider the evolution of two fluids separately, i.e., we can write
\begin{align}
&(\kappa^{2}+\mathcal {F})\bigtriangledown^{\mu}{\sf T}^{\rm (p)}_{\mu \nu}+
\frac{1}{2}\mathcal {F}^{\rm (p)}\bigtriangledown_{\mu}{\sf T}^{\rm (p)}+{\sf T}^{\rm (p)}_{\mu \nu}\bigtriangledown^{\mu}\mathcal {F}^{\rm (p)}=0,\label{relation-1}\\
&\bigtriangledown^{\mu}{\sf T}^{{\rm (u)}}_{\mu \nu}=0,\label{relation-2}
\end{align}
where, $p^{\rm (p)}=0$ has been used. From equation (\ref{relation-2}) we can deduce that in the minimal form of $f({\sf R},{\sf T})$ gravity, the conservation of {\sf EMT} for ultra relativistic fluid, i.e., equation (\ref{ucon}) can be always assumed, at least, as long as mutual interactions are not taken into account. Therefore, regarding the choice (\ref{minimal}) for $f({\sf R},{\sf T})$ function, we can rewrite equation (\ref{relation-1}) as
\begin{align}\label{relation-3}
\Big{(}\kappa^{2} -\frac{3\beta}{2}\Lambda'+\beta \Lambda''{\sf T}\Big{)}\dot{{\sf T}}+3H{\sf T}\Big{(}\kappa^{2} -\beta \Lambda'\Big{)}=0,
\end{align}
where we have used $\rho^{\rm (p)}=-{\sf T}$. Once the function $\Lambda (-{\sf T})$ is determined, the dependency of $-{\sf T}$ and thus $\rho^{({\rm p})}$ on the scale factor can be calculated. More precisely, equation (\ref{relation-3}) can be simplified as
\begin{align}\label{relation-4}
\int_{{\sf T}_{0}}^{-{\sf T}} \frac{\kappa^{2}-\frac{3\beta}{2}\Lambda'+ \beta \Lambda''{\sf T}}{{\sf T}\left(\kappa^{2} -\beta \Lambda'\right)}dT=-3\int_{a_{0}}^{a} d(\ln a),
\end{align}
where, ${\sf T}_{0}$ and $a_{0}$ denote the present values for the scale factor and {\sf EMT} trace. Note that, in general, the integral on the left hand side of equation (\ref{relation-4}) may not be simply solved. Moreover, after the integration process, it may not be possible to clearly write the density as a function of scale factor. Let us choose the functionality of $\Lambda$ parameter as $\Lambda (-{\sf T})=\chi^{2} (-{\sf T})^{\alpha}$, whence we obtain
\begin{align}\label{sol-noncon}
\left[\frac{\left(\rho -\alpha  \beta  \rho ^{\alpha }\right)^{2 \alpha -1}}{\rho }\right]^{\frac{1}{2 (\alpha -1)}}=Ca^{-3},
\end{align}
where $C$ is a constant of integration. In this case, for later application, let us rewrite equation (\ref{relation-3})  in terms of the pressure-less fluid density in the following form
\begin{align}\label{relation-5}
\dot{\rho}^{\rm{(p)}}=-3H\frac{1 -\alpha  \beta  {\rho^{\rm{(p)}}} ^{\alpha-1}}{1-\alpha  \left(\alpha +\frac{1}{2}\right) \beta  {\rho^{\rm{(p)}}}^{\alpha -1}}\rho^{\rm{(p)}},
\end{align}
or correspondingly, eliminating time gives
\begin{align}\label{relation-6}
\left[1-\alpha  (\alpha +\frac{1}{2}) \beta  {\rho^{\rm{(p)}}}^{\alpha -1}\right] \frac{d\rho^{\rm{(p)}}(a)}{da}+3\left(1-\alpha\beta  {\rho^{\rm{(p)}}} ^{\alpha-1}\right)\frac{\rho^{\rm{(p)}}(a)}{a}=0.
\end{align}
In section \ref{sec-conserved} we present an overview on cosmological implications of the only conserved models, i.e., the models with $\Lambda (-{\sf T})=\kappa^{2}\sqrt{-{\sf T}}$. The dynamical system representation of this case has been considered in~\cite{fRT10}. However in this section we review the corresponding cosmological consequences of this case to complete our discussion. Moreover, we will present new details that have not been considered before. In section \ref{sec-nconserved-tot} we consider cosmological behavior of models of type $f({\sf R},{\sf T})={\sf R}+\beta \Lambda (-{\sf T})$ for a general $\Lambda (-{\sf T})$ function via the dynamical system approach. We will see that relaxing the {\sf EMT} conservation which gives equation (\ref{relation-3}), leads to some interesting features; {\sf DE} solutions will be achieved. In section \ref{sec-nconserved} we study the models with $\Lambda (-{\sf T})=\chi^{2} (-{\sf T})^{\alpha}$ when the {\sf EMT} conservation has not been considered. Among non-conserved models there are two special cases that the equivalent dynamical system cannot be constructed properly. More precisely, as we will see the process of recasting the field equations into equivalent dynamical system would breaks down for cases with $\alpha=1$ and $\alpha=-1/2$. In subsections \ref{alpha=1} and \ref{alpha=-1/2} we will consider these cases algebraically.
\/*
 In \{Table (\ref{tab1}) we calculated equation (\ref{sol-noncon}) for three different values of $n$ and also for special case $n=1/2$ (which represent the solution (\ref{specific}, which belongs to the conserved case) and also plotted the diagrams for the evolution of the matter density in the cold {\sf DM} era.
\begin{center}
\begin{table}[h]
\centering
\caption{The evolution of pressure-less matter density implied by equation (\ref{sol-noncon}) for four different values of $n$}
\hskip 0.1in
\begin{minipage}{12cm}
\begin{tabular}{c @{\hskip 0.2in} l@{\hskip 0.15in} l @{\hskip 0.15in} l @{\hskip 0.15in}}\hline
$n$     &$|{\sf T}^{\rm{p}}|(a) =\rho^{\rm{p}}(a)$    &$|{\sf T}^{\rm{p}}|(t) =\rho^{\rm{p}}(t)\footnote{Here, we use the solution $a(t)\propto t^{2/3}$ for the period of pressure-less fluid domination embeded in a geometrically flat Universe.}$ \\[0.5 ex]
\hline
$\frac{1}{2}\footnote{This case represent the only conserved minimal $f({\sf R},{\sf T})$ gravity.}$    &$a^{-3}$ &$t^{-2}$\\[0.75 ex]
$-\frac{1}{2}$    &$\sqrt[3]{\frac{1}{a^9}-\frac{1}{a^{9/2}}+\frac{1}{4}}$&$\sqrt[3]{\frac{1}{t^6}-\frac{1}{t^3}+\frac{1}{4}}$  &\\[0.75 ex]
$1$   &$a^{-12/5}$  &$t^{-8/5}$\\[0.75 ex]
$\frac{3}{2}$  &$\frac{2}{9} \left(-\sqrt{\frac{6}{a^{3/2}}+1}+\frac{3}{a^{3/2}}+1\right)$  &$\frac{2}{9} \left(-\sqrt{\frac{6}{t}+1}+\frac{3}{t}+1\right)$\\[0.75 ex]
\hline
\end{tabular}
\end{minipage}
\label{tab1}
\end{table}
\end{center}

*/
\section{Conserved $\Lambda({\sf T}) {\sf CDM}$ model in phase space}\label{sec-conserved}
\begin{figure}[h]
\centering
\centerline{\epsfig{figure=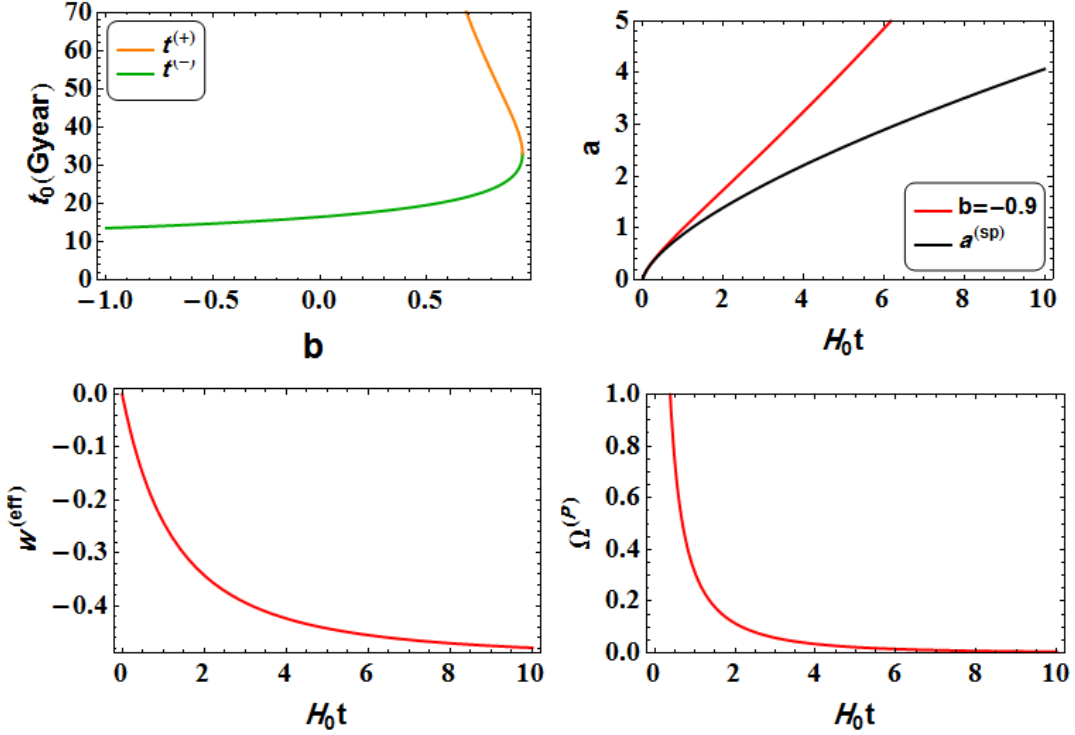,width=15cm}}
\caption{Cosmological quantities for the model $f({\sf R},{\sf T})={\sf R}+\beta \kappa^{2}\sqrt{-{\sf T}}$ when only pressure-less 
fluid is present. In the above plots we have used $b\equiv \beta/H_{0}$. Upper left panel: Two 
solutions for the age of Universe. Lower left panel: the effective {\sf EoS} parameter for $b=-0.9$ 
which is denoted by a red line. Upper right panel: the scale factor of Universe for the 
same value of $b$ ($a^{\rm{sp}}$ shows the scale factor of 
standard matter era). Lower right panel: The behavior of matter density parameter for the same value of $b$.}
\label{fig1}
\end{figure}

In this section we present a brief review on the cosmological solutions of the only case that respect the {\sf EMT} conservation. The {\sf EMT} conservation leads to the constraint equation (\ref{constraint2}) which gives the expression (\ref{specific}) for $\Lambda$ parameter, as the only solution. To illustrate late time effects of the extra term $\sqrt{-{\sf T}}$, we put aside the ultra relativistic fluid. Therefore, field equations (\ref{first}) and (\ref{second}) for the model $f({\sf R},{\sf T})={\sf R}+\beta\kappa^{2}\sqrt{-{\sf T}}$ will take the following form
\begin{align}
&3H^{2}=\kappa^{2}\left(\rho^{\rm{(p)}}-\beta{\rho^{\rm{(p)}}}^{\frac{1}{2}}\right),\label{coneq-1}\\
&2\dot{H}=\kappa^{2}\left(-\rho^{\rm{(p)}}+\frac{\beta}{2}{\rho^{\rm{(p)}}}^{\frac{1}{2}}\right).\label{coneq-2}
\end{align}
The energy conservation yields solution (\ref{pcon}) for which the solution is given by $\rho^{\rm{(p)}}=\rho^{\rm{(p)}}_{0}a^{-3}$, where we have set $\rho^{\rm{(p)}}(a=1)\equiv \rho^{\rm{(p)}}_{0}$. We can check that differentiating equation (\ref{coneq-1}) with respect to time together with using $\dot{\rho}^{\rm{(p)}}=-3H\rho^{\rm{(p)}}$ gives equation (\ref{coneq-2}). This means that solutions of equation (\ref{coneq-1}) would satisfy equation (\ref{coneq-2}). Equation (\ref{coneq-1}) can then be solved to give
\begin{align}\label{coneq-3}
a_{\alpha=\frac{1}{2}}(t)=\left(\frac{3 }{256}\right)^{1/3}\left[\sqrt{\Omega_{0}^{\rm{(p)}}}H_0 t \left(8 \sqrt{3}-3 \beta t\right)\right]^{2/3},
\end{align}
where we have set $\kappa^{2}=1$ and $a(t=0)=0$. We can also check that solution (\ref{coneq-3}) reduces to 
the standard matter dominated era solution for $\beta=0$ and $\Omega_{0}^{\rm{(p)}}=1$. From solution (\ref{coneq-3}), the age of Universe can be calculated as
\begin{align}\label{coneq-4}
t_{0}^{\rm{(\pm)}}=\frac{4 }{\sqrt{3} \beta}\left[1\pm \left(1-\frac{\beta}{\sqrt{3\Omega_{0}^{\rm{(p)}}}H_0}\right)^{\frac{1}{2}}\right],
\end{align}
where $t_{0}^{\rm{(+)}}$ shows the solution with positive sign and $t_{0}^{\rm{(-)}}$ the solution with negative sign between the two terms in brackets. Suppose that $\beta=b H_{0}$ where $b$ is a constant, therefore we have
\begin{align}\label{coneq-5}
t_{0}^{\rm{(\pm)}}=\frac{4\times 9.8 }{\sqrt{3} b H_{0}}\left[1\pm \left(1-\frac{b}{\sqrt{3\Omega_{0}^{\rm{(p)}}}}\right)^{\frac{1}{2}}\right]~~\mbox{Gyear}.
\end{align}
In the {\sf MG} theories we can define an effective EoS parameter as $w^{\rm{(eff)}}\equiv-1-2\dot{H}/3H^{2}$. For the conserved $\Lambda({\sf T}) {\sf CDM}$ model using equations  (\ref{coneq-1}) and  (\ref{coneq-2}) along with solution (\ref{coneq-3}), leads to the following solution for the effective {\sf EoS}
\begin{align}\label{coneq-6}
w^{\rm{(eff)}}=-\frac{1}{2}+\frac{8}{ b \text{t} \left(3 b \text{t}-8 \sqrt{3}\right)+16}.
\end{align}
As can be seen, this solution goes to zero for early times and to $-1/2$ in the late times. Besides, we 
can obtain the fluid density for this case using $\rho^{\rm{(p)}}=\rho^{\rm{(p)}}_{0}a^{-3}$, however in 
this case the scale factor follows the form given in (\ref{coneq-3}). In Figure~\ref{fig1}, we have drawn the 
related plots for cosmological parameters discussed above. The upper left plot presents the
age of Universe for both positive and negative solutions obtained in (\ref{coneq-5}). In this plot the orange line 
denotes the age of Universe for $t_{0}^{\rm{(+)}}$ and the green one indicates $t_{0}^{\rm{(-)}}$. We have 
plotted the rest of diagrams for $b=\pm0.9$. For the purpose of comparison, we have employed a black line for the scale factor of the standard pressure-less fluid dominated era, i.e. $a^{({\rm sp})}$, for which $a(t)\sim t^{2/3}$ holds.

\begin{figure}[ht]
\centering
\centerline{\epsfig{figure=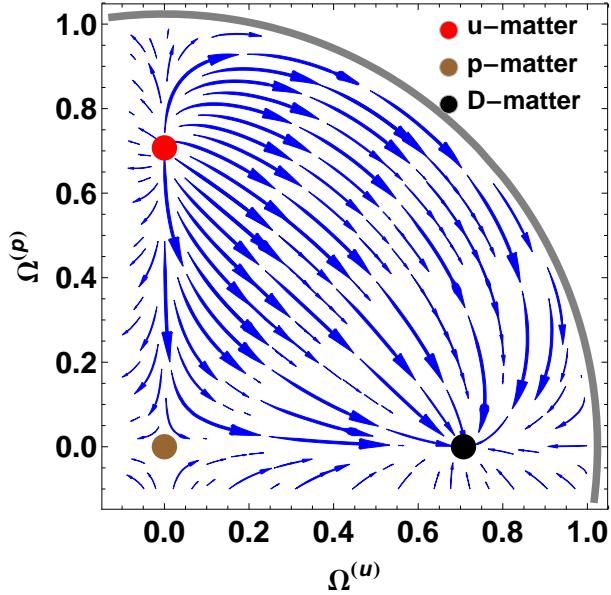,width=8cm}}
\caption{Phase portrait of $f({\sf R},{\sf T})={\sf R}+\beta\kappa^{2}\sqrt{-{\sf T}}$ gravity. The abbreviations \lq\lq{}u-matter\rq\rq{}, \lq\lq{}p-matter\rq\rq{} and \lq\lq{}{\sf DM}\rq\rq{} have been used for specifying the ultra relativistic, pressure-less and {\sf DE} dominated eras, respectively.}
\label{fig2}
\end{figure}

For reconstructing the dynamical system representation of equations  (\ref{coneq-1}) and  (\ref{coneq-2}) we can use the dimensionless variables (\ref{varx4})-(\ref{omega mat}) (remember that in this case we have $x_{5}=x_{4}$/2). In terms of these variables we obtain
\begin{align}
&\Omega^{(\rm{p})}+\Omega^{(\rm{u})}+\Omega^{(\rm{{\sf DE}})}=1,\label{coneq-7}\\
&\frac{d \Omega^{(\rm{u})}}{dt}=\Omega^{(\rm{u})} \left(-\frac{3 \Omega^{(\sf{DE})}}{2}+\Omega^{(\rm{u})}-1\right),\label{coneq-8}\\
&\frac{d \Omega^{(\rm{{\sf DE}})}}{dt}=\Omega^{(\rm{{\sf DE}})}\left[\frac{3 (1-\Omega^{(\rm{{\sf DE}})})}{2}+\Omega^{(\rm{u})}\right],\label{coneq-9}\\
\end{align}
where we have redefined $\Omega^{(\rm{{\sf DE}})}\equiv-\beta x_{4}$. Note that since $x_{4}$ is always positive (see definition (\ref{varx4})), it restricts the allowed values of $\beta$ to negative values, in order that $\Omega^{(\rm{{\sf DE}})}$ stays positive. The fixed points of this dynamical system are presented in Table~\ref{tab1}. Also, we have drawn in Figure~\ref{fig2}, the phase portrait of this model in $(\Omega^{\rm{(u)}},\Omega^{\rm{({\sf DE})}})$ plane.

\begin{center}
\begin{table}[h]
\centering
\caption{The fixed point solutions of $f({\sf R},{\sf T})={\sf R}+\beta\kappa^{2}\sqrt{-{\sf T}}$ gravity.}
\begin{tabular}{l @{\hskip 0.1in} l@{\hskip 0.1in} l @{\hskip 0.1in}l @{\hskip 0.1in}l}\hline\hline\\[-2 ex]
Fixed point     &Coordinates $(\Omega^{\rm{(u)}},\Omega^{\rm{({\sf DE})}})$&Eigenvalues&$w^{\textrm{(eff)}}$\\[0.4 ex]
\hline\\[-2 ex]
$P^{\textrm{({\sf DE})}}$&$(0,1)$&$(-\frac{5}{2},-\frac{3}{2})$&$-\frac{1}{2}$\\[0.75 ex]
$P^{\textrm{(rad)}}$&$(1,0)$&$(\frac{5}{2},1)$&$\frac{1}{3}$\\[0.75 ex]
$P^{\textrm{(m)}}$&$(0,0)$&$(\frac{3}{2},-1)$&$0$\\[0.75 ex]
\hline\hline
\end{tabular}
\label{tab1}
\end{table}
\end{center}
\section{Late time cosmological solutions of $f({\sf R},{\sf T})={\sf R}+\beta \Lambda (-{\sf T})$ gravity}\label{sec-nconserved-tot}
In this section we investigate the late time cosmological solutions of a specific class of models of the form $f({\sf R},{\sf T})={\sf R}+\Lambda (-{\sf T})$. These family of models, can be interesting as they make a minor modification to {\sf GR}. In fact we can interpret these type of models as a $\Lambda({\sf T})${\sf CDM} theory, which imply a matter density dependent cosmological constant. Our aim here is to consider late time solutions only, hence we do not include the ultra relativistic fluid. To reconstruct equations (\ref{eom1}) and (\ref{eom2}) in 
terms of a closed dynamical system for $g({\sf R})={\sf R}$, we use definitions (\ref{varx4}),  (\ref{varx5}), (\ref{n}) and (\ref{parameters}). 
The last two ones, have the role of a (kind of) parametrization\footnote{These types of parametrization have been utilized firstly in 
\cite{amend}.} in determining the functionality of $\Lambda (-{\sf T})$. This parametrization can be suitable 
at least for some well-defined models. Generally, eliminating the trace ${\sf T}$ between definitions (\ref{n}) and (\ref{parameters}) leaves us with a relation between $n$ and $s$ parameters which results in a function $n(s)$. In fact, each $f({\sf R},{\sf T})={\sf R}+\beta \Lambda (-{\sf T})$ model can be 
specified by a function $n(s)$. Models with constant $s$ and $n$ shall be considered in Section \ref{sec-nconserved}. Equations (\ref{eom1}), (\ref{eom2}) and (\ref{relation-3}) can be rewritten in terms of the 
dimensionless variables as follows
\begin{align}
&\Omega^{(\rm{p})}-\beta x_{5}-\frac{\beta}{2}x_{4}=1,\label{total-1}\\
&\frac{2\dot{H}}{3H^{2}}=-1-\frac{\beta}{2}x_{4},\label{total-2}\\
&\frac{\dot{\rho}^{(\rm{p})}}{3H\rho^{(\rm{p})}}=-\frac{1+\frac{\beta}{2}x_{4}}
{1+\frac{\beta}{2}x_{4}-\beta(n+\frac{1}{2})x_{5}}.\label{total-3}
\end{align}
From the definition for effective {\sf EoS} parameter and  equation (\ref{total-2}) we get
\begin{align}
w^{(\rm{eff})}=\frac{\beta}{2}x_{4}.\label{total- 4}
\end{align}
Using equations (\ref{total-1})-(\ref{total-3}) the following autonomous differential equations can be obtained
\begin{align}
&\frac{dx_{4}}{dN}=-3x_{5}\frac{1+\frac{\beta}{2}x_{4}} {1+\frac{\beta}{2}x_{4}-
\beta\left(n+\frac{1}{2}\right)x_{5}}+3x_{4}\left(1+\frac{\beta}{2}x_{4}\right),\label{total- 5}\\
&\frac{dx_{5}}{dN}=-3x_{5}\left(1+\frac{\beta}{2}x_{4}\right)\frac{n-\beta\left(\frac{x_{4}}{2}-\left(n+\frac{1}{2}\right)x_{5}\right)}
{1+\frac{\beta}{2}x_{4}-\beta\left(n+\frac{1}{2}\right)x_{5}}.\label{total- 6}
\end{align}
The above system admits three fixed points with the properties we have listed in Table~\ref{tot}.
\begin{center}
\begin{table}[h]
\centering
\caption{The fixed point solutions of $f({\sf R},{\sf T})={\sf R}+\beta \Lambda (-{\sf T})$ gravity.}
\begin{tabular}{l @{\hskip 0.1in} l@{\hskip 0.1in} l @{\hskip 0.1in}l @{\hskip 0.1in}l@{\hskip 0.1in}l@{\hskip 0.1in}l}\hline\hline\\[-2 ex]
Fixed point     &Coordinates $(x_{4},x_{5})$&Eigenvalues&$w^{\textrm{(eff)}}$&$\Omega^{(\rm{p})}$&$\Omega^{(\rm{{\sf DE}})}$\\[0.4 ex]
\hline\\[-2 ex]
$P^{\textrm{({\sf DE})}}_{1}$&$\left(-\frac{2}{\beta(2n+3)},-\frac{2 (n+1)}{\beta (2 n+3)}\right)$&$\left(\frac{6 n}{2 n+3},-\frac{6 (n+1) \left(n'-1\right)}{2 n+3}\right)$&$-\frac{1}{2n+3}$&$0$&$1$\\[0.75 ex]
$P^{\textrm{({\sf DE})}}_{2}$&$\left(-\frac{2}{\beta},x_{5}\right)$&$\left(0,-\frac{6n}{2n+1}\right)$&$-1$&$-2(1+n)$&$3+2n$\\[0.75 ex]
$P^{\textrm{({\sf DM})}}$&$(0,0)$&$\left(3-3(n+1)n',-3n\right)$&$0$&$1$&$0$\\[0.75 ex]
\hline\hline
\end{tabular}
\label{tot}
\end{table}
\end{center}
Table~\ref{tot} shows that its elements depend on the value of parameter $n$, generally. Note that the {\sf DE} 
density parameter is defined so that the relation $\Omega^{(\rm{p})}+\Omega^{(\rm{{\sf DE}})}=1$ holds and $n'\equiv dn/ds$ 
at the fixed point. From table~\ref{tot} we observe that there is two type of {\sf DE} solutions; $P^{\textrm{({\sf DE})}}_{1}$ which its effective {\sf EoS} parameter depends on $n$ and $P^{\textrm{({\sf DE})}}_{2}$ for which $\Omega^{(\rm{p})}$ 
and thus $\Omega^{(\rm{{\sf DE}})}$ also depend on this parameter. Therefore, only some specific models can give rise to an accelerated expansion solution in the late times. Especially, to be consistent with the observational measurements, the value 
of $n$ parameter can be much more confined. One of the eigenvalues of $P^{\textrm{({\sf DE})}}_{2}$ is zero which shows 
that it is a non-hyperbolic critical point and its stability properties cannot be determined by linear approximation 
techniques. Hence, we focus on the solution characterizing the fixed point $P^{\textrm{({\sf DE})}}_{1}$ .

The fixed points shown in Table~\ref{tot} are solutions of the system $dx_{4}/dN=0,~dx_{5}/dN=0$. 
From this fact we can conclude that for any arbitrary function, namely, $f(x_{4},x_{5})$ we must have 
$df(x_{4},x_{5})/dN=0$ at the equilibrium points. Hence, for parameter $s=s(x_{4},x_{5})$ we obtain
\begin{align}\label{total- 7}
\frac{ds}{dN}=3s(s-n(s)-1)=0.
\end{align}
Therefore, all solutions originated from the presence of function $\Lambda (-{\sf T})$ must satisfy the conditions $s=0$ or $n(s)=s-1$. As can be seen, the latter condition holds for the fixed points $P^{\textrm{({\sf DE})}}_{1}$ (for which we have $x_{5}/x_{4}=n+1=s$) and for the point $P^{\textrm{({\sf DE})}}_{2}$, the constraint equation $x_{5}=sx_{4}=-2(n+1)/\beta$ must hold. The matter fixed point is related to the geometrical sector of the Lagrangian, i.e., ${\sf R}$ and therefore it is not necessary to satisfy the conditions $s=0$ or $n(s)=s-1$. Specifying a function $\Lambda (-{\sf T})$ with the corresponding $n$ parameter, the condition $n(s)=s-1$ leads to an algebraic equation with possible $s_{i}$ roots\footnote{In fact, solutions are intersection points of $n(s)$ curve with the line $s-1$.}. Among them, we only pick out those which exhibit {\sf DE} properties. Briefly speaking, there may be some $s_{i}$ solutions for which, the {\sf DE} fixed points (or at least one) may pass necessary conditions so that a stable late time solution for the underlying model could be achieved. We note that the functions $n(s)$ and $n'(s)$ are calculated for these $s_{i}$ solutions. True cosmological solutions are those that include an unstable {\sf DM} fixed point which is connected to a stable {\sf DE} one. Therefore, discarding the fixed point $P^{\textrm{({\sf DE})}}_{2}$ (owing to vanishing eigenvalue), we seek for conditions on the stability of other points. The stability properties of $P^{\textrm{({\sf DE})}}_{1}$ and $P^{\textrm{({\sf DM})}}$ are determined as follows
\begin{align}
&P^{\textrm{({\sf DE})}}_{1}~~~\mbox{is stable provided} ~~~
\left\{
  \begin{array}{ll}
    \mbox{for}~\hbox{$n'<1$} & \hbox{$-\frac{3}{2}<n<-1$},\\
    \mbox{and for}~\hbox{$n>1$} & \hbox{$-1<n<0$},
  \end{array}
\right.\label{total-8}\\
&P^{\textrm{({\sf DM})}}~~~\mbox{is unstable provided}~~~
\left\{
  \begin{array}{ll}
    \mbox{for}~\hbox{$n'<1$} & \hbox{$n<\frac{1-n'}{n'}$},\\
    \mbox{and for}~\hbox{$n>1$} & \hbox{$n<0$}.
  \end{array}
\right.\label{total-9}
\end{align}
As a result, in order to have the allowed {\sf DM} and {\sf DE} solutions, it suffices that conditions (\ref{total-8}) be satisfied. To complete this section, we explore the discussed method for two specific models. In the next section, we discuss the cosmological solutions for models with a power-law $\Lambda (-{\sf T})$ function.
\begin{itemize}
\item Models with $f({\sf R},{\sf T})={\sf R}+a(-{\sf T})^{\alpha }+b(-{\sf T})^{-\beta }$ where $\alpha>0,~\beta>0$.\\
In this case we obtain
\begin{align}
&n(s)=\alpha -\beta\left(1-\frac{\alpha }{s}\right)-1,\label{total-10}\\
&n'(s)=-\frac{\alpha  \beta }{s^2},\label{total-11}
\end{align}
where, equation $n(s)=s-1$ for (\ref{total-10}) gives rise to the solutions $s_{1}=\alpha$ and $s_{2}=-\beta$. Therefore, a true cosmological solution can be achieved provided that the following conditions hold
\begin{align}
&n'(s_{1})=-\frac{\beta }{\alpha }>1,~~~-\frac{1}{2}<\alpha <0,\label{total-12}\\
&n'(s_{1})=-\frac{\beta }{\alpha }<1,~~~0<\alpha <1,\label{total-13}\\
&n'(s_{2})=-\frac{\alpha }{\beta }>1,~~~0<\beta <\frac{1}{2},\label{total-14}\\
&n'(s_{2})=-\frac{\alpha }{\beta }<1,~~~-1<\beta <0.\label{total-15}
\end{align}
\item Models with $f({\sf R},{\sf T})={\sf R}+a{(-{\sf T})}^{\alpha } \exp^{b(-{\sf T})^{\gamma }}$ for which we get 
\begin{align}
&n(s)=\gamma\left(1 -\frac{\alpha }{s}\right)+s-1,\label{total-16}\\
&n'(s)=\frac{\alpha  \gamma }{s^2}+1.\label{total-17}
\end{align}
The only solution for  equation $n(s)=s-1$ is $s_{\ast}=\alpha$ which leads to $n'(s_{\ast})=1 +\gamma/\alpha$. Hence, the conditions for a true cosmological solution read
\begin{align}
&n'(s_{\ast})=\frac{\alpha +\gamma }{\alpha }>1,~~~ -\frac{1}{2}<\alpha <0,\label{total-18}\\
&n'(s_{\ast})=\frac{\alpha +\gamma }{\alpha }<1,~~~ 0<\alpha <1.\label{total-19}
\end{align}
\end{itemize}
\section{Late time solutions for models with power law $\Lambda (-{\sf T})$  function}\label{sec-nconserved}
In this section, we consider a class of models of type $f({\sf R},{\sf T})={\sf R}+\beta \kappa^2 (-{\sf T})^{\alpha}$ which violate the {\sf EMT} conservation. We will investigate these type of models as dynamical systems as before. We then study, via considering the equilibrium points, cosmological solutions provided by these models. However, for two values of $\alpha$, the dynamical system approach does not work, thereby encouraging us to study these specific cases algebraically. In subsections \ref{alpha=1} and \ref{alpha=-1/2} we will study these specific cases and in subsection \ref{general} we deal with the general form of $f({\sf R},{\sf T})={\sf R}+\beta \kappa^2 (-{\sf T})^{\alpha}$ gravity.
\subsection{Models of type $f({\sf R},{\sf T})={\sf R}+\beta \kappa^2 (-{\sf T})^{\alpha}$ with $\alpha=1$}\label{alpha=1}
For $\alpha=1$, equation~(\ref{relation-6}) reduces to the following equation
\begin{align}\label{alpha1-1}
\left(1-\frac{3 \beta}{2}\right) \frac{d\rho^{\rm{(p)}} (a)}{da}+\frac{3 (1-\beta) \rho^{\rm{(p)}} (a)}{a}=0,
\end{align}
for which the solution is given by
\begin{align}\label{alpha1-2}
\rho ^{\rm{(p)}}(a)=\rho^{\rm{(p)}}_{0} a^{\frac{6 (\beta-1)}{2-3\beta}}.
\end{align}
For $\beta=0$, the above solution leads to the standard form for {\sf DM} energy density and behaves as $a^{2}$ for large values of $\beta$ parameter. In case in which $\beta=1$ we have $\rho^{\rm{(p)}} (a)=\rho^{\rm{(p)}}_{0}$. Moreover for $\alpha=1$, modified Friedman equations (\ref{first}) and (\ref{second}) lead to
\begin{align}
&3H^{2}=\kappa^{2} \left(1-\frac{3 \beta}{2}\right) \rho^{\rm{(p)}},\label{alpha1-3}\\
&2\dot{H}=\kappa^{2}(\beta-1) \rho^{\rm{(p)}}.\label{alpha1-4}
\end{align}
Inserting solution~(\ref{alpha1-2}) into equation (\ref{alpha1-3}) and solving the resultant equation, we get the scale factor as a function of time, as
\begin{align}\label{alpha1-5}
a_{\alpha=1} (t)=\left(\frac{3}{2}\right)^{\frac{2-3 \beta }{6(1- \beta)}} \left[\frac{3(1-\beta ) \sqrt{\Omega_{0}^{p}} H_{0}t}{\sqrt{2-3 \beta }}\right]^{\frac{2-3 \beta }{3 (1-\beta )}}.
\end{align}
This solution is valid only for $\beta<2/3$ and leads to the standard form for the pressure-less matter dominated era for $\beta=0$. We can also check that solutions (\ref{alpha1-2}) and (\ref{alpha1-5}) satisfy equation (\ref{alpha1-4}). Applying the definition given for effective {\sf EoS} on equations  (\ref{alpha1-3}) and  (\ref{alpha1-4}) we see that these type of models correspond a constant value $w^{\rm{(eff)}}=\beta/(2-3\beta)$, which for $\beta<2/3$ gives $w^{\rm{(eff)}}>-1/3$. However, there is a special case; equations (\ref{alpha1-2}), (\ref{alpha1-3}) and  (\ref{alpha1-4}) yield a de Sitter solution for $\beta=1$. As a result, power law models with $\alpha=1$, result in a single decelerated expanding cosmological state for the Universe (at all times) without ever passing through a pressure-less matter dominated era. However, these models predict a single de Sitter state, as well.
%
\subsection{Models of type $f({\sf R},{\sf T})={\sf R}+\beta \kappa^2 (-{\sf T})^{\alpha}$ with $\alpha=-1/2$}\label{alpha=-1/2}
\begin{figure}[h]
\centering
\centerline{\epsfig{figure=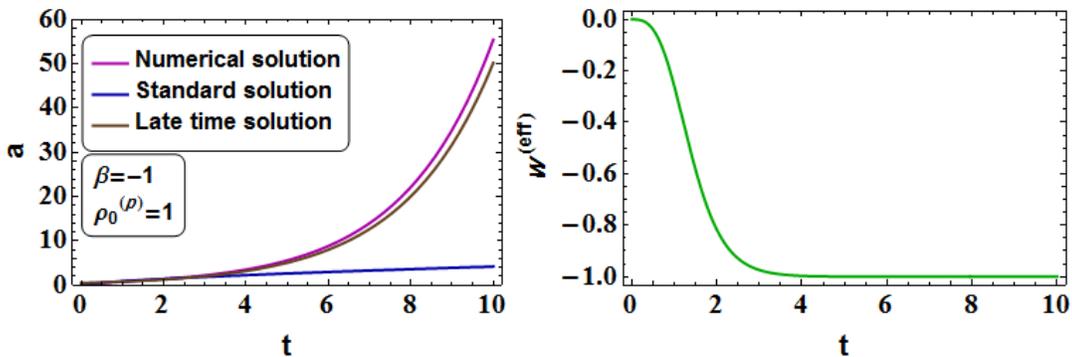,width=15cm}}
\caption{Some cosmological parameters for power law model with $\alpha=-1/2$. Left panel: numerical solution of the scale factor in purple, approximate solution in brown and solution for the standard matter dominated era in blue. Right panel: The evolution of effective {\sf EoS} parameter. }
\label{fig3}
\end{figure}

In this case we have $-\frac{3}{2}\Lambda'+ \Lambda''{\sf T}=0$, thus, equation~(\ref{relation-6}) reduces to the following simple form
\begin{align}\label{alpha minus12-1}
\frac{d\rho^{(\rm{p})}(a)}{da}+3\left(\frac{1}{2}\beta {\rho^{(\rm{p})}}^{-3/2}(a)+1\right)\frac{\rho ^{(\rm{p})}(a)}{a}=0,
\end{align}
which has the following solution for the matter density in terms of the scale factor
\begin{align}\label{alpha minus12-2}
\rho^{(\rm{p})} (a)=2^{-2/3}\left(\frac{\beta+2{\rho^{(\rm{p})}_{0}}^{3/2}}{a^{9/2}}-\beta\right)^{2/3},
\end{align}
where we have set $\rho^{(\rm{p})}(a=1)\equiv\rho_{0}$. This solution in the early times where $a\rightarrow0$ behaves as $a^{-3}$ and in the late times where $a\rightarrow\infty$, as $(-\beta/2)^{2/3}$. The Friedman equations can then be obtained as
\begin{align}
&3H^{2}=\kappa^{2} \rho^{(\rm{p})},\label{alpha minus12-3}\\
&2\dot{H}=-\kappa^{2}\left(\rho^{(\rm{p})}+\frac{\beta}{2}{\rho^{(\rm{p})}}^{-1/2}\right).\label{alpha minus12-4}
\end{align}
As can be seen, the first modified Friedman equation reduces to its standard form (i.e., its form in {\sf GR}), and for this reason writing down the equations as a physically consistent dynamical system breaks down. Exact solutions of equation~(\ref{alpha minus12-3}) with energy density of {\sf DM} given by (\ref{alpha minus12-2}) cannot be obtained, explicitly. Nevertheless, we can obtain the early and late time solutions, being given as
\begin{align}
&a_{\alpha=-1/2}^{(\rm{early})}(t)\approx\left(\frac{27}{256}\right)^{1/9}\left[\beta+2(3H_{0}^{2}\Omega_{0}^{\rm{(p)}})^{3/2}\right]^{2/9}t^{2/3},\label{alpha minus12-5}\\
&a_{\alpha=-1/2}^{(\rm{late})}(t)\approx e^{\frac{1}{6} \sqrt[3]{-\beta} \left(2^{2/3} \sqrt{3} t-\frac{4}{(\beta+2)^{\frac{1}{3}}}\right)}.\label{alpha minus12-6}
\end{align}
From solution~(\ref{alpha minus12-5}) we see that the term in square brackets reduces to the standard form for pressure-less matter i.e., $(3H_{0}^{2}/2)^{2/3}$, once we set $\beta=0$. The above solutions have been obtained so that in the present time they would be equal to unity. Solution (\ref{alpha minus12-6}) shows that in the late times, we have a de Sitter solution for $-2<\beta<0$ (notice that for the late time solution, i.e., $(-\beta/2)^{2/3}$,  equations~(\ref{alpha minus12-3}) and~(\ref{alpha minus12-4}) give $H=constant$ and $\dot{H}=0$, respectively). In the left panel of Figure~\ref{fig3} we have plotted the scale factor for three cases; numerical diagram has been drawn from equations (\ref{alpha minus12-3}) for ~(\ref{alpha minus12-2}) in purple color, asymptotic curve in the late times, i.e. solution (\ref{alpha minus12-6}) in brown color and the scale factor for the standard pressure-less dominated era in blue one, for comparison. We used $\beta=-1$ and $\rho_{0}^{\rm{(p)}}=1$ for plotting purpose. In the right panel of Figure \ref{fig3}, we have presented the effective {\sf EoS} for the same values of parameters $\beta$ and $\rho_{0}^{\rm{(p)}}$. These diagrams show that in the model $f({\sf R},{\sf T})={\sf R}-\kappa^{2} (-{\sf T})^{-1/2}$, the Universe could experience an accelerated expansion in the late times, when only a pressure-less fluid is present. It is interesting to note that, the unusual interaction between the only cosmological fluid and curvature of space-time, which leads to the violation of {\sf EMT} conservation, has the consequence of a late de Sitter epoch in the evolution of the Universe.
\subsection{Models of type: $f({\sf R},{\sf T})={\sf R}+\beta \kappa^2 (-{\sf T})^{\alpha}$ ($\alpha\neq1,-1/2$)}\label{general}
\begin{figure}[t!]
\centering
\centerline{\epsfig{figure=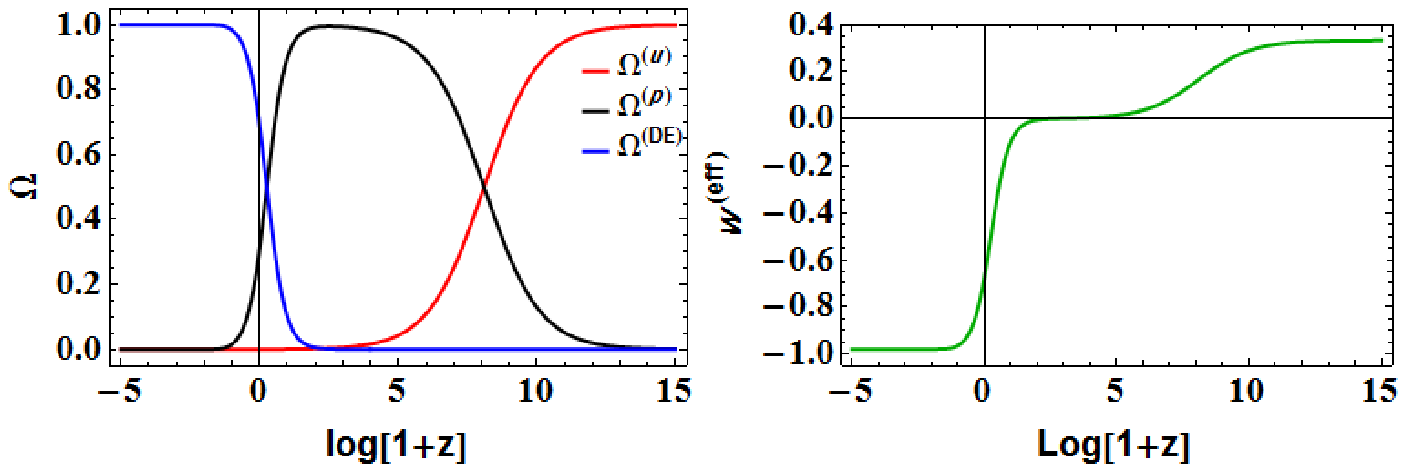,width=16cm}}\vspace{2mm}
\centerline{\epsfig{figure=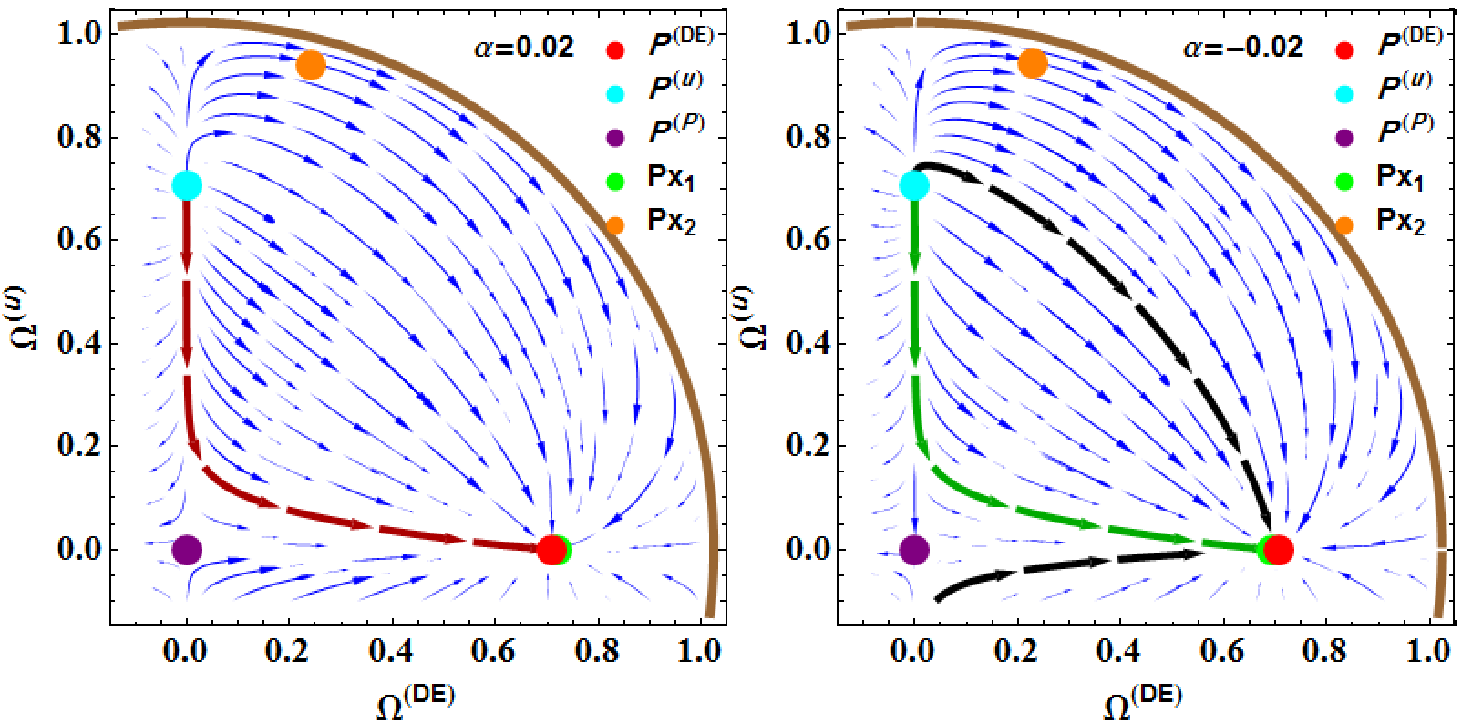,width=16cm}}
\caption{Cosmological quantities for models $f({\sf R},{\sf T})={\sf R}+\beta (-{\sf T})^{\alpha}$. Upper panels: the matter density parameters and the effective {\sf EoS} for $\alpha=0.02$ and initial values $\Omega^{\rm{({\sf DE})}}_{i}=9.7\times10^{-23}$ and $\Omega^{\rm{(u)}}_{i}=0.999$. Lower panels: phase space portraits for two indicated values of $\alpha$. Red and green trajectories show the possible physically justified solutions and black trajectories suffer from either lacking a matter dominated or radiation dominated eras.}
\label{fig4}
\end{figure}

In subsections \ref{alpha=1} and \ref{alpha=-1/2} we have considered cosmological implications of two specific models that could not be analyzed by the dynamical system approach. In the present subsection we will provide a comprehensive study these models via this approach. We will see that unlike the conserved case, there is however, a de Sitter solution in the case of non-conserved models. For those models in which $f({\sf R},{\sf T})={\sf R} + \beta \Lambda (-{\sf T})$, the field equations (\ref{first}) and (\ref{second}) reduce to the following equations
\begin{align}
&3H^{2}=-\kappa^{2}\left({\sf T}^{\textrm{(p)}}-\beta h'{\sf T}^{\textrm{(p)}}+\frac{\beta}{2}h-\rho^{\rm{(u)}}\right),\label{general1}\\
&2\dot{H}=\kappa^{2}\left({\sf T}^{\textrm{(p)}}-\beta h'{\sf T}^{\textrm{(p)}}-\frac{4}{3}\rho^{\textrm{(u)}}\right).\label{general2}
\end{align}
Substituting $\Lambda (-{\sf T})=\kappa^{2} (-{\sf T})^{\alpha}$  into equations (\ref{general1}), (\ref{general2}) together with using (\ref{relation-3}) leads to
\begin{align}
&3H^{2}=\kappa^{2}\left[\rho^{\textrm{(p)}}-\beta\left(\alpha+\frac{1}{2}\right){\rho^{\rm{(p)}}}^{\alpha}+\rho^{\rm{(u)}}\right],\label{general3}\\
&2\dot{H}=\kappa^{2}\left(-\rho^{\textrm{(p)}}+\beta\alpha{\rho^{\rm{(p)}}}^{\alpha}-\frac{4}{3}\rho^{\textrm{(u)}}\right),\label{general4}\\
&\dot{\rho}^{\rm{(p)}}=-3H\frac{1-\beta\alpha{\rho^{\rm{(p)}}}^{\alpha-1}}{1-\beta\alpha\left(\alpha+\frac{1}{2}\right){\rho^{\rm{(p)}}}^{\alpha-1}}\rho^{\textrm{(p)}}.\label{general5}
\end{align}
As can be seen, equations (\ref{general3}) and (\ref{general4}) reduce to equations (\ref{alpha1-3}) and (\ref{alpha1-4}) for $\alpha=1$ and in the case of $\alpha=-1/2$, these equations reduce to (\ref{alpha minus12-3}) and (\ref{alpha minus12-4}) in the absence of ultra relativistic fluid. Using definitions (\ref{varx4}), (\ref{omega rad}) and (\ref{omega mat}), the above equations can be rewritten as  
\begin{align}
&\Omega^{\rm{(p)}}-\beta\left(\alpha+\frac{1}{2}\right)x_{4}+\Omega^{\rm{(u)}}=1,\label{general6}\\
&\frac{2\dot{H}}{3H^{2}}=-1-\frac{\beta}{2}x_{4}-\frac{1}{3}\Omega^{\rm{(u)}},\label{general7}\\
&\dot{\rho}^{\rm{(p)}}=-3H\frac{1+\frac{\beta}{2}x_{4}}{1+\beta(1-\alpha)\left(\alpha+\frac{1}{2}\right)x_{4}}\rho^{\rm{(p)}}.\label{general8}
\end{align}
Finally, the dimensionless evolutionary equations for variables $\Omega^{\rm{(u)}}$ and $x_{4}$ are obtained as
\begin{align}
&\frac{d\Omega^{\rm{(u)}}}{dN}=\Omega^{\rm{(u)}}\left(-1+\frac{3\beta}{2}x_{4}+\Omega^{\rm{(u)}}\right),\label{general9}\\
&\frac{dx_{4}}{dN}=x_{4}\left\{\frac{3(1-\alpha)\left(1+\frac{\beta}{2}x_{4}\right)\left[1+\beta\left(\alpha+\frac{1}{2}\right)x_{4}\right]}{1+\beta(1-\alpha)\left(\alpha+\frac{1}{2}\right)x_{4}}+ \Omega^{\rm{(u)}}\right\}.\label{general10}
\end{align}
Note that in the case of the power law dependency, we have $x_{5}=x_{4}/2$ which demands a slightly 
deferent system of equations with respect to a general $\Lambda (-{\sf T})$ function. As can be seen, equation (\ref{general10}) implies no {\sf DE} solution for $\alpha=1$. We cannot also interpret a {\sf DE} model from it for $\alpha=-1/2$, since equation (\ref{general6}) does not include a {\sf DE} term. Moreover, one can check that equations  (\ref{general6}),  (\ref{general9}) and  (\ref{general10}) reduce to equations (\ref{coneq-7}), (\ref{coneq-8}) and (\ref{coneq-9}) for $\alpha=1/2$ and $\beta=-1$. Also, we can obtain the effective {\sf EoS} from equation (\ref{general7}) as follows
\begin{align}\label{general11}
w^{\rm{(eff)}}=\frac{\beta}{2}x_{4}+\frac{1}{3}\Omega^{\rm{(u)}}.
\end{align}
The fixed point solutions of the system (\ref{general9}) and  (\ref{general10}) are calculated in Table~\ref{tab2}.
\begin{center}
\begin{table}[h]
\centering
\caption{The fixed point solutions for $f({\sf R},{\sf T})={\sf R}+\beta\kappa^{2} (-{\sf T})^{\alpha}$ gravity with radiation.}
\begin{tabular}{l @{\hskip 0.1in} l@{\hskip 0.1in} l @{\hskip 0.1in}l @{\hskip 0.1in}l@{\hskip 0.1in}l@{\hskip 0.1in}l}\hline\hline\\[-2 ex]
Fixed point     &Coordinates $(\Omega^{\rm{(u)}},x_{4})$&Eigenvalues&$w^{\textrm{(eff)}}$&$\Omega^{\rm{(p)}}$&$\Omega^{\rm{({\sf DE})}}$\\[0.4 ex]
\hline\\[-2 ex]
$P^{\textrm{({\sf DE})}}$&$\left(0,-\frac{1}{\beta\left( \alpha+\frac{1}{2}\right) }\right)$&$\left(3-\frac{9}{2 \alpha +1},-\frac{2 (\alpha +2)}{2 \alpha +1}\right)$&$-\frac{1}{2 \alpha +1}$&$0$&$1$\\[0.75 ex]
$P^{\textrm{(u)}}$&$(1,0)$&$(4-3 \alpha,1)$&$\frac{1}{3}$&$0$&$0$\\[0.75 ex]
$P^{\textrm{(p)}}$&$(0,0)$&$\Big(3(1-\alpha),-1\Big)$&$0$&$1$&$0$\\[0.75 ex]
$Px_{1}$&$\left(0,-\frac{2}{\beta }\right)$&$\left(\frac{6 (1-\alpha)}{2 \alpha -1},-4\right)$&$-1$&$-2 \alpha$&$1+2\alpha$\\[0.75 ex]
$Px_{2}$&$\left(\frac{8 \left(\alpha ^2+\alpha -2\right)}{\alpha  (8 \alpha -1)-4},-\frac{8-6 \alpha }{\left(-8 \alpha ^2+\alpha +4\right) \beta }\right)$&$\Big(f(\alpha),g(\alpha)\Big)$&$\frac{1}{3}$&$i(\alpha)$&$j(\alpha)$\\[0.75 ex]
\hline\hline
\end{tabular}
\label{tab2}
\end{table}
\end{center}
Defining $\Omega^{\rm{({\sf DE})}}\equiv -\beta\left(\alpha+\frac{1}{2}\right)x_{4}$, provided that $\Omega^{\rm{(p)}}+\Omega^{\rm{({\sf DE})}}+\Omega^{\rm{(u)}}=1$, the fixed point $P^{\textrm{({\sf DE})}}$ can be accounted as a {\sf DE} solution for which we have $\Omega^{\rm{({\sf DE})}}=1$. The condition for accelerated expansion, $w^{\textrm{(eff)}}<-1/3$ is satisfied only for $-\frac{1}{2}<\alpha <1$.
It is interesting to note that only for $-\frac{1}{2}<\alpha <1$ both eigenvalues become negative, simultaneously. Therefore, in non-conserved class of models of the type $f({\sf R},{\sf T})={\sf R}+\beta \kappa^2 (-{\sf T})^{\alpha}$, there is a stable solution for late times with the following properties
\begin{align}\label{general12}
P^{\textrm{({\sf DE})}}=\left(\Omega^{\rm{(u)}}=0, \Omega^{\rm{(p)}}=0, \Omega^{\rm{({\sf DE})}}=1\right),~~w^{\textrm{(eff)}}<-1/3,~~\mbox{for}~~~-\frac{1}{2}<\alpha <1.
\end{align}
It is noteworthy that, for $\alpha\rightarrow 0^{\pm}$, we have $w_{{\rm eff}}\rightarrow -1$ so that we can get a {\sf DE} solution with observationally accepted values of the {\sf EoS} parameter. Planck 2015 measurements show that the Universe is undergoing an accelerated expansion driven by {\sf DE} which its present values of the {\sf EoS} parameter lie in the interval $-1.051<w^{\rm{{\sf DE}}}_{0}<-0.961$~\cite{Planck}. This fact imposes a constraint on the $\alpha$ parameter as, $-0.024\lesssim\alpha\lesssim0.020$. In addition to the existence of a {\sf DE} fixed point, there are unstable solutions which indicate domination of the ultra relativistic and pressure-less fluids, i.e., the points $P^{\textrm{(u)}}$ and $P^{\textrm{(p)}}$, respectively. Two other solutions are $Px_{1}$ and $Px_{2}$ which are not physically interesting, since they do not correspond to dominant cosmological epochs\footnote{The related values for density parameters of pressure-less fluid and {\sf DE} are given by, $i(\alpha)=-\frac{2 (\alpha +2) (3 \alpha -4)}{\alpha  (8 \alpha -1)-4}$ and $j(\alpha)=-\frac{8-6 \alpha }{\left(-8 \alpha ^2+\alpha +4\right) \beta }$, respectively.}. Nevertheless, $Px_{1}$ can be a late time solution for small values of $\alpha$. Table~\ref{tab2} shows that the fixed point $Px_{1}$  is a stable one for small values of $\alpha$. Therefore, depending on the value of $\alpha$, each of these solutions may be the late time solution. To show the two possibilities, we have plotted in Figure~\ref{fig4}, the phase space diagrams for the values $\alpha=0.02$ and $\alpha=-0.02$. The red solid circle denotes fixed point $P^{\textrm{({\sf DE})}}$, the green one indicates $Px_{1}$, the purple solid circle shows $P^{\rm{(p)}}$, the cyan one indicates $P^{\rm{(u)}}$ and the orange solid circle shows $Px_{2}$. Diagrams show that physically interesting trajectories begin from $P^{\rm{(u)}}$, pass along $P^{\rm{(p)}}$ and then terminate at either $P^{\rm{({\sf DE})}}$ or $Px_{1}$. Figure~\ref{fig4} also shows that for $\alpha<0$, the fixed point $Px_{1}$ is the late time solutions and for $\alpha>0 $ the solution $P^{\rm{({\sf DE})}}$ would be chosen. These points will overlap for $\alpha=0$, that is the {\sf GR} model plus a cosmological constant, known as the celebrated $\Lambda {\sf CDM}$ model. The fixed point $Px_{2}$ coincides with the fixed point $P^{\textrm{(u)}}$ for $\alpha=4/3$, otherwise it is physically meaningless. The fixed point $Px_{2}$ is always unstable, that is, the functions $f(\alpha)$ and $g(\alpha)$ never get the same negative sign or pure imaginary values. In Figure~\ref{fig4} diagrams for evolution of the matter density parameters and the effective {\sf EoS} parameter are plotted for $\alpha=0.02$, as well.
\section{Conclusion}\label{conclusion}
In this work we have investigated the cosmological consequences of violation of {\sf EMT} conservation for a class of $f({\sf R},{\sf T})$ theories of gravity. We have considered both the ultra relativistic fluid and {\sf DM} in a spatially flat, homogeneous and isotropic background given by the {\sf FLRW} metric. We have studied models of type $f({\sf R},{\sf T})={\sf R}+\beta \Lambda (-{\sf T})$ which we call these as minimally coupled $f({\sf R},{\sf T})$ models. This specific model can be considered as a $\Lambda({\sf T}){\sf CDM}$ model which allows a density dependency for the cosmological constant. Firstly, we have presented the field equations of $f({\sf R},{\sf T})$ gravity and defined some dimensionless variables. We also classified the minimal models to those that respect the conservation of {\sf EMT} and those that do not. The former models have been considered elsewhere, however to complete our study, we have briefly reviewed their cosmological solutions through the dynamical system approach. Some new results have been obtained as well. We have algebraically showed that these type of models cannot be accepted since they have a late time solution with an undesirable {\sf EoS} parameter. Their {\sf EoS} parameter varies from zero to $-1/2$ which is not observationally confirmed. Thus, considering the {\sf EMT} conservation, {\sf GR} theory modified by a minimal $\Lambda(-{\sf T})$ function has still the problem of incompatibility with recent observational outcomes.
The latter models do not respect the conservation of {\sf EMT}, for which a modified version of {\sf DM} density conservation have been obtained. We have shown that in the minimal models of $f({\sf R},{\sf T})$ gravity, it is always possible to consider the evolution of ultra relativistic fluid and {\sf DM} independent of each other, as long as interactions are turned off. Therefore, only a modification in the behavior of {\sf DM} density can provide a different cosmological scenario, at least in the late time epochs.

To consider the cosmological consequences of the violation of {\sf EMT} conservation, we presented a general method to formulate the dynamical system equations for generic minimally coupled models. We have defined two dimensionless $n$ and $s$ parameters constructed out of $\Lambda (-{\sf T})$ function and its derivatives and showed that the resulted autonomous equations will depend upon these parameters. As a result, we have obtained a set of closed equations which their solutions, and hence the stability properties of them, will be controlled by these parameters. We have discussed that at least for well behavior models, we can parametrize $\Lambda (-{\sf T})$ function in terms of a function $n(s)$. We illustrated that, all fixed points originated from $\Lambda (-{\sf T})$ function, must lie on the line $n=s-1$. In other words, for every function $n(s)$ all fixed point solutions must occur at the location where the $n(s)$ curve intersects with the line $n=s-1$. The fixed points representing the ultra relativistic and pressure-less matter domination, are not subjected to the above discussion as they are solutions of {\sf GR}. Briefly speaking, this method shows that, there generally exist two fixed point solutions which represent accelerated expansion/{\sf DE} era in the late times. We have applied the method to inspect two models specified by $f({\sf R},{\sf T})={\sf R}+a(-{\sf T})^{\alpha }+b(-{\sf T})^{-\beta }$ and $f({\sf R},{\sf T})={\sf R}+a{(-{\sf T})}^{\alpha } \exp^{b(-{\sf T})^{\gamma }}$ and discussed the validity of their solutions.

As a special case that cannot be explained by the above technique, we investigated the cosmological behavior of models with $\Lambda (-{\sf T})=\kappa^{2} (-{\sf T})^{\alpha}$. In this case, we included the ultra relativistic fluid and discussed the late time solutions. We found that there are two different {\sf DE} solutions which their properties depend on the constant $\alpha$. Nevertheless, each $f({\sf R},{\sf T})={\sf R}+\beta \Lambda (-{\sf T})$ model accepts only one solution, which accordingly, such type of $f({\sf R},{\sf T})$ functions can be classified as two different models. We have argued that observationally consistent models may be constructed by small values of $\alpha$. For example, Planck 2015 measurements have shown that if we believe in {\sf DE} as one of the ingredients of the Universe which is presently driving the observed accelerated expansion, its {\sf EoS} parameter must lie within $-1.051<w_{0}^{({\sf DE})}<-0.961$. This fact restricts us to accept $-0.024<\alpha<0.02$. We have shown that for two specific models with $\alpha=1$ and $\alpha=1/2$,  the dynamical system approach does not work, due to the of structure of the related equations. Algebraic treatments showed that the former model, in general, indicates a single decelerating late time cosmological era. However, there is an exact single de Sitter solution, as well. For the determined values of $\alpha$ and $\beta$ parameters, we obtained a constant value for the {\sf EoS} parameter with a specific value lying within the range $w_{\alpha=1}>-1/3$. The latter case gives a proper solution including a connected matter and {\sf DE} dominated eras. This model accepts a de Sitter solution in the late times. Finally, we would like to end this article by highlighting the importance of examining the observable signals of {\sf MG} theories, in order to test the physical validity of the resulted models. If experiments confirm that a modified version of {\sf GR} can explain observations better than the original version, the results could shed light on some fundamental cosmological questions. Modified gravity theories have been utilized successfully to account for galaxy cluster masses \cite{galclmass}, the velocity field of {\sf DM} and galaxies \cite{veldmgalprl}, the cosmic shear data \cite{cossheardata}, the rotation curves of galaxies \cite{rotcgal}, velocity dispersions of satellite galaxies \cite{veldissaga}, and globular clusters \cite{gloclust}. These theories have been also used to propose an explanation for the Bullet Cluster \cite{BullClus} without resorting to nonbaryonic {\sf DM}, see also \cite{MOTO} and references therein. However, among the {\sf MG} theories that have been proposed so far, Rastall\rq{}s gravity touches one of the cornerstones of {\sf GR}, i.e., the conservation of {\sf EMT} \cite{Rastall} and interestingly, this issue has been entered within the context of {\sf MG} theories \cite{rasmodgr}. While in the present work, we have studied cosmological consequences of violation of {\sf EMT} in the framework of $f({\sf R},{\sf T})$ gravity theory, it is of utmost importance to seek for observational evidences (such as gold sample supernova type Ia data \cite{gssuperIA}, SNLS supernova type Ia data set \cite{SNLSdata} and X-ray galaxy clusters analysis \cite{XrayClus}) that could distinguish between the resulted models from this theory and {\sf GR}. However, observational features of this theory needs to be carried out with more scrutiny and dealing with this issue is beyond the scope of the present paper.
%

\end{document}